\documentclass[aps,prb,twocolumn,noshowpacs]{revtex4-1}
\usepackage{amsmath,graphicx,amsbsy,amssymb,amsmath,epsfig,latexsym,wasysym,pifont,mathrsfs,natbib}
\usepackage{float} \newfloat{widefig}{thp}{lop} \usepackage{comment}
\usepackage{array, makecell} \usepackage{verbatim} \usepackage{datetime}
\usepackage{multirow,array} \usepackage{hyperref} \usepackage{color} \usepackage{setspace}
\def\boxit#1{%
  \smash{\color{black}\fboxrule=1pt\relax\fboxsep=0pt\llap{\rlap{\fbox{\strut\makebox[#1]{}}}~}}\ignorespaces
}

\begin{document}

\title{Giant magnetocaloric effect driven by first-order magneto-structural transition in cosubstituted Ni-Mn-Sb Heusler compounds: predictions from \textit{Ab initio} and Monte Carlo calculations}
\author{Sheuly Ghosh}
\email[]{sheuly.ghosh@iitg.ac.in}
\affiliation{Department of Physics, Indian Institute of Technology
  Guwahati, Guwahati-781039, Assam, India.}    
\author{Subhradip Ghosh}
\email{subhra@iitg.ac.in} \affiliation{Department of Physics,
  Indian Institute of Technology Guwahati, Guwahati-781039, Assam,
  India.} 

\begin{abstract}
Using Density Functional Theory and a thermodynamic model [Physical Review B 86, 134418 (2012)], in this paper, we provide an approach to systematically screen compounds of a given Heusler family to predict ones that can yield giant magnetocaloric effect driven by a first-order magneto-structural transition. We apply this approach to two Heusler series Ni$_{2-x}$Fe$_{x}$Mn$_{1+z-y}$Cu$_{y}$Sb$_{1-z}$ and Ni$_{2-x}$Co$_{x}$Mn$_{1+z-y}$Cu$_{y}$Sb$_{1-z}$, obtained by cosubstitution at Ni and Mn sites. We predict four new compounds with potentials to achieve the target properties. Our computations of the thermodynamic parameters, relevant for magnetocaloric applications, show that the improvement in the parameters in the predicted cosubstituted compounds can be as large as four times in comparison to the off-stoichiometric Ni-Mn-Sb and a compound derived by single substitution at the Ni site, where magnetocaloric effects have been observed experimentally. This work establishes a protocol to select new compounds that can exhibit large magnetocaloric effects and demonstrate cosubstitution as a route for more flexible tuneability to achieve outcomes, better than the existing ones.
\end{abstract}

\maketitle

\section{Introduction}\label{introduction}

The development of magnetic refrigeration, a new solid-state refrigeration technology, based on the magnetocaloric effect (MCE), continues to attract considerable attention worldwide due to its environmentally friendly nature, higher energy efficiency, lower mechanical noise, and simple mechanical construction, in comparison with conventional technology based on gas compression/expansion\cite{oliva1989,tegusi2003,banks2002}. The underlying magnetocaloric effect (MCE) is measured in terms of isothermal magnetic entropy change ($\Delta$S$_\mathrm{mag}$) and/or adiabatic temperature change ($\Delta$T$_\mathrm{ad}$) that require large variations in the $\text{material}$'s magnetization with temperatures. In the magnetic refrigerators, Gd has been considered as a benchmark material due to the discovery of significant magnetocaloric effect in it, an outcome of a second-order ferromagnetic to paramagnetic transition close to room temperature\cite{hughes2007}. However giant effect is generally observed in materials which undergo a first-order magneto-structural transition i.e. a structural phase transition, coupled with a magnetic one\cite{pecharsky1997,pecharsky2001,tishin2016,de2006,fujieda2002,wada2001,tegus2002,li2012,li2014,krenke2005,7krenke2007,8pasquale2005,pathak2007,79krenke2005,muthu2010}. Magnetic refrigeration near room temperature is of special interest because of its social and economic benefits. From this point of view, the continuous search of new solid-state magnetic refrigerants that could exhibit a giant MCE, in an appropriate temperature range, as well as the improvement of the existing ones, have been focus of research in this area. \\

Among MCE materials, shape memory Heusler compounds are of great interest as they exhibit large MCE, and their transition temperatures can be easily tuned. The origin of their large MCE is the first-order martensitic phase transition (MPT) from a high temperature cubic austenite phase to a low temperature low symmetry phase, a large magnetization change occurring simultaneously. One of the compounds in the Heusler family which recently showed promising MCE is the off-stoichiometric Mn-excess Sb-deficient Ni-Mn-Sb where magneto-structural transition and significant magnetocaloric effect were observed near room temperature\cite{91duc2012,22khan2007,nayak2009}.

With an aim to improve the MCE in this family of compounds, some recent investigations have also been carried out, by substituting the Fe and Co atoms at Mn and Ni sites, respectively, resulting in large positive values of $\Delta$S$_\mathrm{mag}$ \cite{nayak2009,anayak2009,64han2008,59sahoo2011}. Though the transition metal substituted Ni-Mn-Sb Heusler compounds turn out to be useful materials exhibiting giant MCE, one major disadvantage is that upon substitution, the working temperature, i.e., the martensitic transformation temperature (T$_{M}$) falls below the room temperature. This is not desirable for operational purposes. In some recent studies, the strategy of substitution of at least two $3d$~elements simultaneously (cosubstitution) has been found to be useful in achieving the important magnetic and structural properties with better tuning and adaptability\cite{zeleny2014,55sokolovskiy2015,perez2018}. In a recent work\cite{ghosh2020co}, we explored potential room-temperature magnetocaloric materials in two cosubstituted families, Ni$_{2-x}$Fe$_{x}$Mn$_{1+z-y}$Cu$_{y}$Sb$_{1-z}$ (Fe@Ni-Cu@Mn) and Ni$_{2-x}$Co$_{x}$Mn$_{1+z-y}$Cu$_{y}$Sb$_{1-z}$ (Co@Ni-Cu@Mn). We found that for a certain range of compositions, there is no structural transformation down to the low temperature indicating that the MCE is purely due to second-order magnetic phase transition. We also found that large magnetic moments and T$_{c}$, the Curie temperature, close to room temperatures can be easily achieved by tuning the compositions, along with a significant MCE. These indicated a delicate balance of the concentrations of different constituents and easy tunability of properties in this family. Armed with this information, in the present work, we focused on the composition ranges that were not covered in Ref ~\onlinecite{ghosh2020co}. A martensitic phase transformation (MPT) occurs in cosubstituted Ni-Mn-Sb compounds with those compositions. We aimed to explore whether a giant MCE due to magneto-structural coupling can be predicted in these compositions along with near room temperature T$_{M}$. Using a thermodynamic model in conjunction with first-principles electronic structure calculations, we made comparisons with the systems already explored experimentally and provided predictions of new compounds, yet to be realized experimentally, that can exhibit significantly large MCE. This study established a systematic way to use the information on structural and magnetic properties obtained from first-principles calculations to screen the materials that are potential ones with target properties and the robustness of the formalism to accurate predictions of new compounds.

\section{Method of Calculation and Computational Details}\label{methods}
Most of the literature on MCE, particularly for Heusler compounds, is experimental in nature. Very few theoretical studies of MCE in the framework of molecular-field approximation\cite{amaral2007,alvarez2011,szalowski2011}, bond proportion model\cite{triguero2006}, and Monte Carlo simulations\cite{nobrega2006,1buchelnikov2010,buchelnikov2011,buchelnikov2010,nobrega2005} have been reported. Tackling the problem using theoretical tools is a difficult one as the task to obtain all phases in a self-consistent way is quite demanding. This requires the equilibrium, {\it ab initio} evaluation of all magnetic exchange parameters and comparison of free energies, for each structure (austenite and martensite) at different temperatures. Avoiding all the above mentioned complex and time-consuming calculations, a unified description of structural and associated first-order magnetic phase transition has been presented in literature, successfully for Ni-Mn based Heusler compounds, by using a simple model Hamiltonian consisting of tuneable parameters\cite{1buchelnikov2010,buchelnikov2011,buchelnikov2010,singh2013,1sokolovskiy2013,sokolovskiy2013}. The Hamiltonian allowed one to explore the richness of the phase diagram. The observed qualitative and quantitative behavior of MCE quantities turned out to be in very good agreement with experiments. Therefore, we have adopted the same method in this work.

\subsection{First-principles methods and computational details}\label{1-method}
We have used the first-principles electronic structure calculations to gain information on the phase stability of the compounds explored and their magnetic properties. 
The electronic structure calculations were done with spin-polarised density functional theory (DFT) based projector augmented wave (PAW) method as implemented in Vienna {\it Ab initio} Simulation Package (VASP)\cite{41blochl1994,43kresse1999,42kresse1996}. The valence electronic configurations used for the Mn, Fe, Co, Ni, Cu and Sb PAW pseudopotentials are 3$d^{6}$4$s$, 3$d^{7}$4$s$, 3$d^{8}$4$s$, 3$d^{8}$4$s^{2}$, 3$d^{10}$4$s$ and 5$s^{2}$5$p^{3}$, respectively. For all calculations, we used the Perdew-Burke-Ernzerhof implementation of generalized gradient approximation for exchange-correlation functional\cite{44perdew1996}. An energy cut off of 550 eV, and a Monkhorst-Pack $11\times{11}\times{11}$ k-mesh were used for self-consistent calculations. The convergence criteria for the total energies and the forces on individual atoms were set to 10$^{-6}$ eV and $10^{-2}$ eV/\r{A} respectively. 

The stabilities of the compounds against decomposition into its components were checked by computing the formation energies:
\begin{eqnarray}
E_{f}=E_{tot}-\sum_{i}n_{i}E_{i}
\end{eqnarray}

$E_{tot}$ is the total electronic energy of the systems, $i$ represents the atoms in the unit cell, and $n_{i}$ is the concentration of the $i$-th atom. $E_{i}$ is the total energy of the element $i$ in its bulk ground state.

To compute the Curie temperature T$_{c}$ of a compound, we first calculated the magnetic pair exchange parameters using multiple scattering Green function formalism (KKR) as implemented in SPRKKR code\cite{ebert2011}. In here, the spin part of the Hamiltonian was mapped to a Heisenberg model,

\begin{eqnarray}
H = -\sum_{\mu,\nu}\sum_{i,j}
J^{\mu\nu}_{ij}
\mathbf{e}^{\mu}_{i}
.\mathbf{e}^{\nu}_{j}
\end{eqnarray}

$\mu$, $\nu$ represent different sub-lattices, \emph{i}, \emph{j} represent atomic positions and $\mathbf{e}^{\mu}_{i}$ denotes the unit vector along the direction of magnetic moments at site \emph{i} belonging to sub-lattice $\mu$. The $J^{\mu \nu}_{ij}$s were calculated from the energy differences due to infinitesimally small orientations of a pair of spins within the formulation of Liechtenstein {\it et al.}\cite{liechtenstein1987}. In order to calculate the energy differences by the SPRKKR code, full potential spin-polarized scalar relativistic Hamiltonian with angular momentum cut-off $l_{max} = 3$ was used along with a converged k-mesh for Brillouin zone integrations. The Green's functions were calculated for 32 complex energy points distributed on a semi-circular contour. The energy convergence criterion was set to 10$^{-5}$ eV for the self-consistent cycles. These exchange parameters were then used for the calculation of T$_c$. The Curie temperatures were estimated with two different approaches: the mean-field approximation (MFA)\cite{26sokolovskiy2012,meinert2010} and the Monte Carlo simulation (MCS) method\cite{landau2014,zagrebin2016,28bkundu2017}. Details of the calculations using these methods are given in the supplementary material. \\
 
\subsection{Calculation of MCE parameters using thermodynamic model}\label{2-method}
The Monte Carlo simulation is an appropriate tool to use the zero-temperature \textit{ab initio} calculations for computing properties at finite temperatures. Here, we have used Monte Carlo simulation on a model Hamiltonian to estimate the  MCE parameters, i.e., isothermal change in magnetic entropy ($\Delta$S$_\mathrm{mag}$) or adiabatic temperature change ($\Delta$T$_\mathrm{ad}$) due to the application of a magnetic field. The model Hamiltonian was chosen such that it accommodates, along with magnetic and structural degrees of freedom, the coupling between the two\cite{buchelnikov2010,buchelnikov2011,sokolovskiy2013,1sokolovskiy2013}. \\

The model Hamiltonian ($H$) consists of three parts: (a) the magnetic contribution due to the magnetic degrees of freedom of the system, $H_{m}$; (b) the elastic contribution due to the structural transformation from cubic to tetragonal phases, $H_{els}$; and (c) the contribution arising from the coupling of magnetic and structural interactions, $H_{int}$. \\

\begin{eqnarray}
H=H_{m}+H_{els}+H_{int} 
\label{eq-all}
\end{eqnarray}

The magnetic subsystem is described by a mixed q-states Potts model\cite{buchelnikov2010,singh2013,sokolovskiy2013,meyer2000,singh2011}, which allows for both first- and second-order phase transitions, where q is the number of spin states for magnetic atoms.
\begin{eqnarray}
H_{m}  = -\sum_{<i,\:j>}^{NN} J_{ij} \delta_{S_i,\:S_j} - g\mu_B\mu_0\mathrm{H}_\mathrm{ext}\:\sum_{i}^{N} \delta_{S_i,\:S_g} 
\label{eq-potts}
\end{eqnarray}

Here, the first term represents the magnetic interactions at different lattice sites; $J_{i,j}$ being the exchange parameters involving sites $i$ and $j$, $S_{i}$ the spin defined on the lattice site $i=1, 2, ....., N$ and $N$ the total number of atoms considered in the simulation cell. The second term represents the coupling of the spin system to the external magnetic field H$_\mathrm{ext}$ along the direction of ghost spin variable $S_{g}$. $\mu_{\rm B}$ is the Bohr magneton, g is the Lande factor (here $g=2$). \\

The degenerate Blume-Emery-Griffiths (BEG) model\cite{blume1971,vives1996}, which allows one to describe the interaction between the elastic variables, was used to address the mutual influence of magnetic ordering and structural transitions. The energy of the system undergoing structural transformation can be represented by,

\begin{eqnarray}
H_{els}  = -J\sum_{<i,\:j>}^{NN} \sigma_i \sigma_j -K\sum_{<i,\:j>}^{NN}(1-\sigma_i^2)(1-\sigma_j^2) \nonumber  \\
-k_{B}T\ln (p)\sum_{i}(1-\sigma_i^2) \nonumber  \\
-K1g\mu_B\mu_0 \mathrm{H}_\mathrm{ext} \sum_{i}^{NN} \delta_{\sigma_g \sigma_i} \sum_{<i,\:j>}^{NN} \sigma_i \sigma_j
\label{eq-struc}
\end{eqnarray}

where $\sigma$ is the strain parameter and denotes the structural state of the lattice site which takes the value 0 for cubic or undistorted state, and +1 or -1 for the tetragonal or distorted
state. $J$ and $K$ are structural exchange constants for tetragonal and cubic states respectively, $p$ is the degeneracy factor characterizing the number of tetragonal states, $K1$ is the dimensionless magneto-elastic interaction, and $T$ is the temperature of the system. The third term accounts for the higher configurational entropy in the cubic phase. The last term accounts for the energy contribution due to the changes in the structural states under the influence of the external magnetic field. The structural states are coupled to the external magnetic field through the ghost spin state $\sigma_{g}$. The sign of the magneto-elastic parameter, $K1$, indicates the favored structural state (cubic or tetragonal), in presence of an external magnetic field. For $K1>0$, energy is removed from the system in a tetragonal state, so that the tetragonal (distorted) state is favored over cubic state by the external magnetic field, while for a $K1<0$, energy is added to the system in a tetragonal state so that the cubic state is favored. In essence, $K1>(<)0$ if T$_M$ increases(decreases) in the presence of external magnetic field. 
 
\begin{eqnarray}
H_{int} = 2\sum_{<i,\:j>}^{NN}U_{ij}\delta_{S_i,\:S_j}(\frac{1}{2}-\sigma_i^2)(\frac{1}{2}-\sigma_j^2) \nonumber  \\
-\frac{1}{2}\sum_{<i,\:j>}^{NN}U_{ij}\delta_{S_i,\:S_j}
\label{eq-interaction}
\end{eqnarray}

In the magneto-elastic part (equation \ref{eq-interaction}) of the Hamiltonian, the first term describes the effective coupling of magnetic sub-lattice to the modulation of the lattice, while the last term renormalizes the spin-spin interaction. $U_{ij}$ is the magneto-elastic interaction parameter. \\

This Hamiltonian (equation \ref{eq-all}) was used for Monte Carlo calculations using the following procedure: \\

1. For all the magnetic and lattice sites in the supercell, values of initial spin ($S_{i}$) and strain ($\sigma$$_{i}$) were chosen as 1. \\

2. First, one arbitrary lattice site was chosen, and the initial elastic energy contribution ($H_{els}^{initial}$) from that site was calculated using equation \ref{eq-struc}. For $\sigma_{i}=0$, the $\mathrm{site}$'s energy contribution was calculated on a cubic lattice, while for $\sigma_{i}=+1$ or $\sigma_{i}=-1$, the energy was calculated on a tetragonal lattice. This was also done while calculating the magnetic and coupled contributions of the Hamiltonian. \\

3. For the same site $i$, the strain parameter $\sigma_{i}$ was changed randomly. The elastic energy of the new configuration, $H_{els}^{final}$, was calculated. The change in the elastic energy of the system, $(H_{els}^{final}-H_{els}^{initial})$, was then computed. \\

4. The new system configuration i.e., the lattice with a new strain parameter on that particular site was accepted or rejected based on the Metropolis algorithm\cite{landau2014,zagrebin2016,newman1999}. If $H_{els}^{final}$ $\leq$ $H_{els}^{initial}$, The new system configuration was accepted; else a ratio $R$ is calculated

\begin{eqnarray} 
R=\exp(-(H_{els}^{final}-H_{els}^{initial})/k_BT) 
\label{eq-ratio}
\end{eqnarray}

A random number ($r$; 0 $<$ r $<$ 1) is generated, and if $R > r$, the new configuration was accepted. \\

5. The next step was to find the new spin state for the site $i$. The total energy of the system, $H^{initial}$, was calculated using equation \ref{eq-all}. The spin state of site $i$ was then changed randomly, and the energy of the new configuration, $H^{final}$, was calculated. The new system configuration (site $i$ with new spin parameter) was then accepted or rejected
based on the Metropolis Algorithm, as described in the previous step. \\

6. The steps (2) - (5) were then repeated by moving through the lattice sites. Once all the lattice sites were swept, one Monte Carlo step (MCS) was completed. \\

At a given temperature, the system was first equilibrated by repeating the above Monte Carlo steps. Then the average magnetization ($m$) and strain order parameter ($\varepsilon$) of the equilibrated system for a given temperature were calculated as,

\begin{eqnarray}
m=\frac{1}{N}\sum_i^n \frac{(q_iN_i^{max}-N_i)}{q_i-1}
\label{mag-order}
\end{eqnarray}

\begin{eqnarray}
\varepsilon=\frac{1}{N}\sum_i\sigma_i
\end{eqnarray}   

Where $i$ in equation (\ref{mag-order}) denotes a magnetic atom type, $n$ the total number of magnetic atom types, $q_i$  the total number of spin states of the atom of type $i$, $N_i^{max}$ is the maximum number of atoms of type $i$ with the same spin state, $N_i$ is the total number of atoms of type $i$, $N$ is the total number of atoms in the system. \\
 The magnetic specific heat ($C_\mathrm{mag}$), the magnetic entropy (S$_\mathrm{mag}$) and total specific heat, $C = C_\mathrm{lat} + C_\mathrm{mag}$ with lattice  and magnetic contributions were then calculated. We have neglected the electronic part of the specific heat. For the lattice heat, we have used the standard Debye approximation. Finally, the magnetocaloric parameters i.e. the isothermal changes in magnetic entropy ($\Delta$S$_\mathrm{mag}$) and the adiabatic temperature change ($\Delta$T$_\mathrm{ad}$) due to the application of an external field, were calculated by equation (\ref{eq-del-smag}) and (\ref{eq-del-Tad}) respectively.

\begin{eqnarray}
C_\mathrm{mag}(T,\mathrm{H}_\mathrm{ext})=\frac{1}{N}\sum_i^N \frac{\langle H^{2} \rangle - \langle H \rangle^{2}}{k_{B}T^2}
\label{eq-cmag}
\end{eqnarray}

\begin{eqnarray}
\mathrm{S}_\mathrm{mag}(T,\mathrm{H}_\mathrm{ext})=\frac{1}{N}\int_{0}^{T}dT\frac{C_\mathrm{mag}(T,\mathrm{H}_\mathrm{ext})}{T}
\label{eq-smag}
\end{eqnarray}

\begin{eqnarray}
C_\mathrm{lat}(T,\Theta_D)=9RN_i\Big\{4\Big(\frac{T}{\Theta_D}\Big)^3\int_{0}^{\Theta_D/T}dx\frac{x^3}{e^x-1} \nonumber  \\
-\Big(\frac{\Theta_D}{T}\Big)\frac{1}{e^{(\Theta_D/T)}-1}\Big\}
\label{eq-clat}
\end{eqnarray}

\begin{eqnarray}
\Delta \mathrm{S}_\mathrm{mag}(T,\mathrm{H}_\mathrm{ext})=\mathrm{S}_\mathrm{mag}(T,\mathrm{H}_\mathrm{ext}) - \mathrm{S}_\mathrm{mag}(T,0)
\label{eq-del-smag}
\end{eqnarray}   

\begin{eqnarray}
\Delta T_\mathrm{ad}(T,\mathrm{H}_\mathrm{ext})=-T\frac{\Delta \mathrm{S}_\mathrm{mag}(T,\mathrm{H}_\mathrm{ext})}{C(T,\mathrm{H}_\mathrm{ext})}
\label{eq-del-Tad}
\end{eqnarray}

\section{Results and Discussions}\label{results}
In this work, we investigated the two cosubstituted Mn-excess, Sb-deficient Ni-Mn-Sb families: (i) Ni$_{2-x}$Fe$_{x}$Mn$_{1+z-y}$Cu$_{y}$Sb$_{1-z}$ (denoted as Fe@Ni-Cu@Mn) and (ii) Ni$_{2-x}$Co$_{x}$Mn$_{1+z-y}$Cu$_{y}$Sb$_{1-z}$ (denoted as Co@Ni-Cu@Mn). As mentioned in the section~\ref{introduction}, our motivation was to find compounds that give rise to large MCE driven by first-order magneto-structural transition. In order to model the compounds that have multi-sublattice chemical disorder, we have considered a 16 atom conventional cubic unit cell. This unit cell mimics the high temperature austenite (L2$_{1}$ Heusler) phase of the systems (space group 225). The consequence of the cell size is the inability to model compositions with arbitrary $x$, $y$, or $z$. Each one of the three variables can be changed by an amount of 0.25 only. The choice of the composition range to achieve ones with the target properties is crucial. For this, we took recourse to our previous works on the same system \cite{ghosh2020co,ghosh2020}. Based upon the findings there, we restricted the value of $x$ to 0.25, and the range of $z$ between 0.50 and 0.75. The variable $y$ is constrained to be less than or equal to the value of $z$. The compounds with these compositions are expected to exhibit a martensitic phase transformation. 

Since, site occupancies in the compounds do not always follow a regular pattern (the excess atoms occupying the sites of deficient atoms in the reference system) \cite{82ghosh2014,35sanchez2007,ghosh2019,45li2011,28bkundu2017,89chakrabarti2013} affecting the MCE properties as a consequence, we first found out the minimum energy configuration for each composition by comparing the total energies of configurations with different site occupancy patterns and magnetic configurations. We found that the substituting Fe/Co atoms prefer to occupy the Ni sites, whereas the Cu atoms prefer to occupy the Sb sites, the same as found out in Ref ~\onlinecite{ghosh2020co}. We also found that depending upon the composition, two types of magnetic configurations are found leading to minimum total energy; FM, where the two types of Mn can align parallel and FIM, where the two types of Mn can align antiparallel. 

After fixing the lowest energy configurations, we first calculated the formation energies for all the compounds. Negative values of the formation energy for each of the composition (Table \ref{table1}) indicated their stability against decomposition into its constituent elements. Subsequently, the following criteria were used to screen materials further to narrow down the ones which are potential large MCE materials:

(i) The materials should possess a high magnetic moment in their austenite phases. \\
(ii) The martensitic transformation temperature (T$_{M}$) should be near room temperature or higher than that. \\
(iii) The MPT should be associated with a substantial change in magnetization. In other words, a magnetic structure in the martensite phase different than that in the austenite phase would be advantageous. \\
(iv) The second-order magnetic transition temperature, i.e., the Curie temperature (T$_c^A$) in the austenite phase, should be close to the martensitic transformation temperature (T$_{M}$). This would lead to a large change in entropy due to near simultaneous magnetic and structural transition. Otherwise, T$_c^A$ should be higher than T$_{M}$ so that the MPT can occur in a magnetically ordered phase. \\

In what follows, we present and analyze the results on the variations in the magnetic moments in the austenite phases, the relative stabilities of the structural phases and variations in the T$_{M}$ and the variations in T$_{c}^{A}$ with changes in the compositions by systematic variations in $x,y$ and $z$. The analysis helps us in the prediction of new compounds that have potentials to exhibit large MCE. In here, we first understand the trends in the physical quantities due to cases with single substitution. For that, we extended the range of $x$ up to $0.75$ in all substituted compounds. The outcome of the cosubstituted cases can be understood in the light of the results for single substituted compounds.

\subsection{Magnetic moment in austenite phase} \label{moment}
\begin{figure}[t]
\centerline{\hfill
\psfig{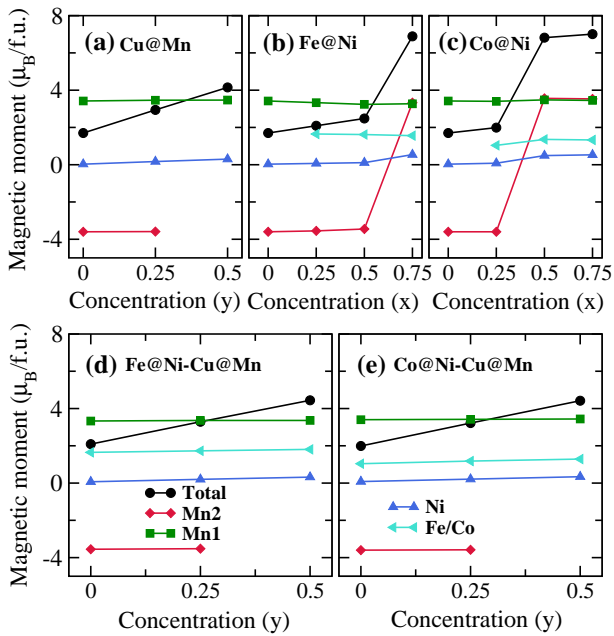}
\hfill}
\caption{Variations in the calculated total and atomic magnetic moments (in $\mu_{\rm B}/f.u.$) with (a) $y$ ($x$=0), ((b)-(c)) $x$ ($y$=0) and ((d)-(e)) $y$ ($x$=0.25) for Ni$_{2-x}$Fe$_{x}$Mn$_{1+z-y}$Cu$_{y}$Sb$_{1-z}$ (Fe@Ni-Cu@Mn) and Ni$_{2-x}$Co$_{x}$Mn$_{1+z-y}$Cu$_{y}$Sb$_{1-z}$ (Co@Ni-Cu@Mn) systems in their austenite phases. $z$ value is kept at $0.5$ throughout. Mn1 and Mn2 denote Mn atoms at its own site and at other sites in L2$_{1}$ structure, respectively.} 
\label{fig1}
\end{figure}

\begin{figure}[t]
\centerline{\hfill
\psfig{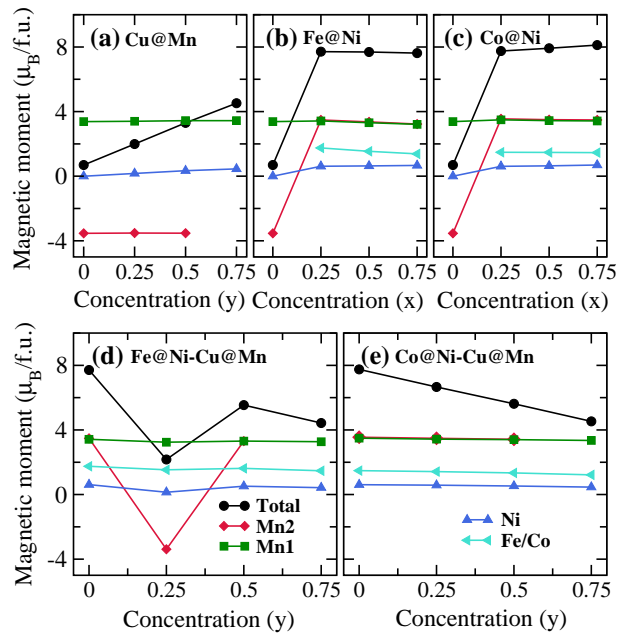}
\hfill}
\caption{Variations in the calculated total and atomic magnetic moments (in $\mu_{\rm B}/f.u.$) with (a) $y$ ($x$=0), ((b)-(c)) $x$ ($y$=0) and ((d)-(e)) $y$ ($x$=0.25) for Ni$_{2-x}$Fe$_{x}$Mn$_{1+z-y}$Cu$_{y}$Sb$_{1-z}$ (Fe@Ni-Cu@Mn) and Ni$_{2-x}$Co$_{x}$Mn$_{1+z-y}$Cu$_{y}$Sb$_{1-z}$ (Co@Ni-Cu@Mn) systems in their austenite phases. $z$ is kept at $0.75$ throughout. Mn1 and Mn2 denote Mn atoms at its own site and at other sites in L2$_{1}$ structure, respectively.} 
\label{fig2}
\end{figure}

A correlation between enhancement of magnetization in the austenite phase and a large MCE for Ni-Mn-Sb system could be observed in the experiments\cite{nayak2009,anayak2009}. For Ni$_{1.8}$Co$_{0.2}$Mn$_{1.52}$Sb$_{0.48}$\cite{nayak2009}, an enhanced magnetic moment in the austenite phase has been observed due to the Co substitution at Ni site. Subsequently, a large positive MCE was observed in the system, presumably an artifact of magneto-elastic coupling. In general, a significant enhancement of magnetic moment in the austenite phase leads to a possibility of large $\Delta$M, the difference in magnetization between austenite and martensitic phases, the key to a first-order magneto-structural transition. We, therefore, focus on finding the possibility of enhancement of magnetic moment in austenite phases of Ni-Mn-Sb compounds upon substitution by $3d$-elements at different sites. We present the results on total and atomic magnetic moments in Fig.~\ref{fig1} and Fig.~\ref{fig2}. The panels (a)-(c) in each of the two figures show results for single substitution while (d)-(e) show results for cosubstitution. All results in Fig.~\ref{fig1} are for compounds with $z=0.5$ that is with 50$\%$ excess Mn (with respect to stoichiometric Ni$_{2}$MnSb) while those in Fig.~\ref{fig2} are with $z=0.75$. If we first look at the compounds with single substitution (Table 1, supplementary material), we find that irrespective of $z$, Cu substitution at the Mn site allows the total moment to increase linearly with the Cu content $y$. This is due to the fact that the atomic moments of both Mn atoms stay nearly same and that the gradual replacement of Mn2 atoms by Cu reinforces the moment since Mn2, being aligned anti-parallel to Mn1, was reducing the total moment. When Fe or Co substitutes Ni, irrespective of the value of $z$, the behavior of magnetic moment with the concentration of Fe or Co, $x$, is qualitatively identical in the sense that a monotonic variation with $x$ is either preceded or followed by a discontinuous jump at a critical value of $x$; the difference being in the critical value. Such discontinuous jump with at least two-fold increase in the total moment occurs due to the change in the magnetic structure from FIM to FM, driven by the orientations of the Mn atoms. The variations in the moments for cosubstitution with $x$ fixed at $0.25$, turn out to be the combined behavior of the two single-substituted cases, Fe or Co substituting Ni and Cu substituting Mn. Due to the presence of higher concentration of Mn2 atoms in compounds with $z=0.75$, in comparison to those with $z=0.5$, the overall moments in the former cosubstituted compounds are higher than that in the later (Table \ref{table1}). The inference from these results is that the compositions with $z=0.75$ may provide higher values of $\Delta$M and thus better MCE than compounds with composition having $z=0.5$, closest to the one on which experiments have been performed.

\begin{figure*}[t]
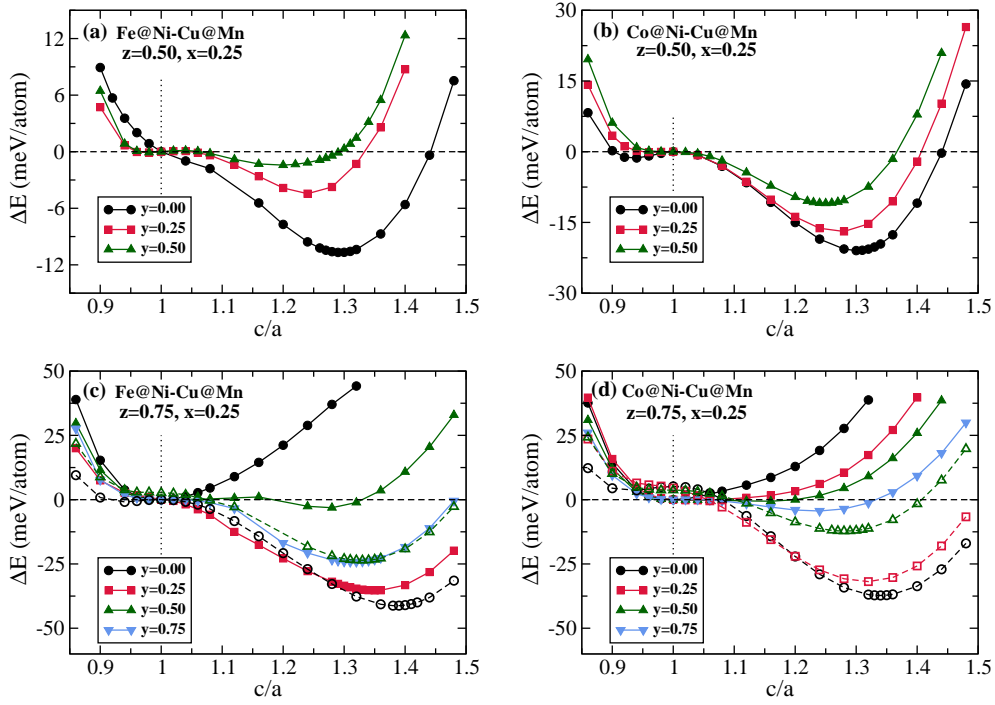

\centerline{\hfill
\psfig{file=fig3a.eps,width=0.35\textwidth}
\hspace{0.3cm}
\psfig{file=fig3b.eps,width=0.35\textwidth}
\hfill}
\vspace{0.3cm}
\centerline{\hfill
\psfig{file=fig3c.eps,width=0.35\textwidth}
\hspace{0.3cm}
\psfig{file=fig3d.eps,width=0.35\textwidth}
\hfill}
\caption{The variations of total energy difference ($\Delta$E) between the austenite(L2$_{1}$) and the martensite(tetragonal) phases as a function of tetragonal distortion i.e. $c/a$ ratio in Ni$_{2-x}$(Fe/Co)$_{x}$Mn$_{1+z-y}$Cu$_{y}$Sb$_{1-z}$ ((Fe/Co)@Ni-Cu@Mn) compounds for different $y$ with ((a)-(b)) $z$=0.50 and ((c)-(d)) $z$=0.75. $x$ is fixed at the value 0.25 for all calculations.The magnetic configurations considered in each case is given in Table \ref{table1}. The solid curves are for cases where magnetic configurations considered are same in both structural phases. The dotted curves in the panels (c) and (d) are for cases where the magnetic configurations in the tetragonal phases are different from those in the Heusler phases and provide minimum total energies.}
\label{fig3}
\end{figure*}
\subsection{Martensitic transformation and magnetic structures across structural phases} \label{E-vs-cbya}
Next, we investigate whether MPT occurs in our systems of investigation. To this end, we distort the austenite L2$_{1}$ structure along the $z$-axis by keeping the volume at the equilibrium value of the austenite phase and compute the total energy of the system as a function of the tetragonal distortion given by $(c/a)$. Typical profiles of compounds undergoing MPT will have their minima at $(c/a)\neq 1$. For all compositions, we calculated the energy profiles as a function of tetragonal distortion. The results are shown in Fig.~ \ref{fig3}. The results suggest that all the considered compounds, undergo MPT, a requirement for further consideration. However, for compounds with higher Mn content ($z=0.75$), the magnetic ground states of the austenite phases are different than those in the martensitic phases except at the compound where $y=0.75$. In all such cases, the austenite phase has FM magnetic structure (Table \ref{table1}), while the martensitic phase has  FIM magnetic structure. The results in Fig.~\ref{fig3}(c)-(d) corroborate this. A consequence of this is a large value of $\Delta$M (Table \ref{table1}) for the compounds with $z=0.75$ as compared to those with $z=0.5$. This makes the compounds with $z=0.75$ potentially better to realize large MCE.  

\begin{figure}[htpb!]
\centerline{\hfill
\psfig{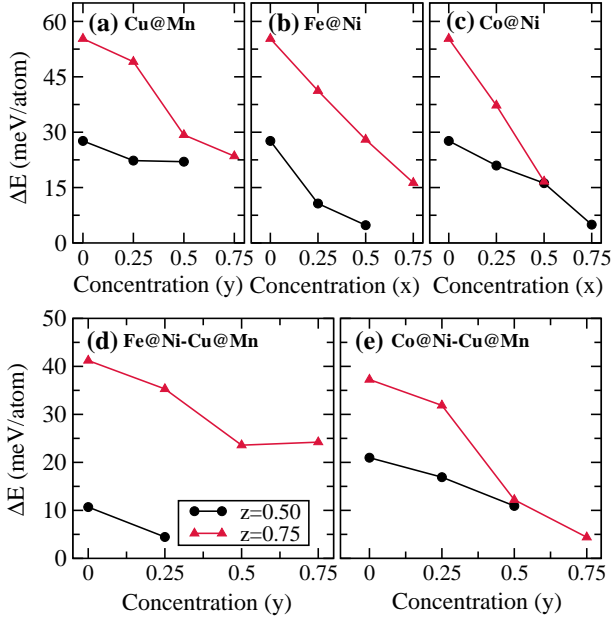}
\hfill}
\caption{Variations in total energy difference, $\Delta$E between the austenite(L2$_{1}$) and the martensite(tetragonal) phases with (a) $y$ ($x$=0), ((b)-(c)) $x$ ($y$=0) and ((d)-(e)) $y$ ($x$=0.25) for Ni$_{2-x}$(Fe/Co)$_{x}$Mn$_{1+z-y}$Cu$_{y}$Sb$_{1-z}$ compounds with different values of $z$.} 
\label{fig4}
\end{figure}

In order to make sure that this is indeed so, we looked at the variations in $\Delta$E, the energy difference between the austenite and martensite phases in their respective ground states. The results for single-substituted compounds are shown in Fig.~\ref{fig4}(a)-(c) while those for cosubstituted ones are shown in Fig.~\ref{fig4}(d)-(e). The $\Delta$E values are also listed in Table \ref{table1}. In literature, $\Delta$E is routinely used to predict the martensitic transformation temperature (T$_{M}$)\cite{89chakrabarti2013,81sokolovskiy2017,28bkundu2017,ghosh2019,ghosh2020}. Here we have used it first to understand the trends in the T$_{M}$ so that compositions with higher T$_{M}$ can be screened. From Fig.~\ref{fig4}, we find that the trends in variations of $\Delta$E with compositions in cases of the cosubstituted compounds can be correlated with the trends in case of single-substituted ones. A general trend of $z=0.75$ compounds having higher $\Delta$E and thus higher T$_{M}$ can be immediately inferred. Therefore, in both the counts of larger $\Delta$M and higher T$_{M}$, the compounds with Mn-content as high as 1.75 can be considered promising to obtain large MCE.

\subsection{Curie temperature in austenite phase} \label{curie-temp}

\begin{figure}[H]
\centerline{\hfill
\psfig{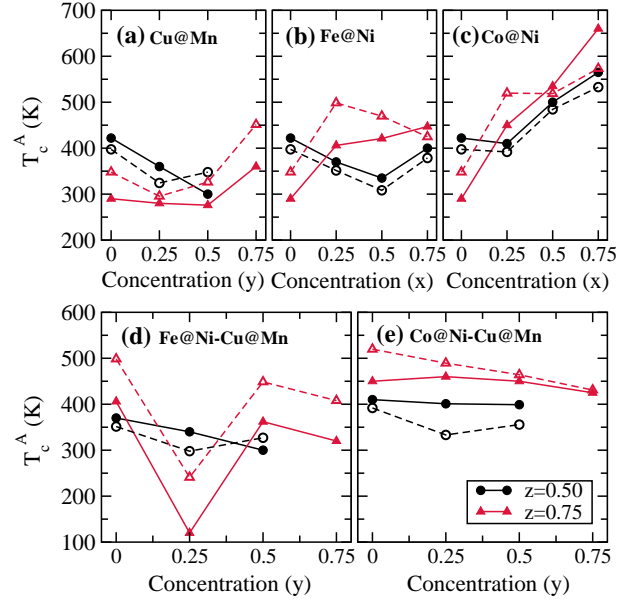}
\hfill}
\caption{Variations in calculated Curie temperatures (T$_c^A$) in austenite phases with concentration (a) $y$ ($x$=0), ((b)-(c)) $x$ ($y$=0) and ((d)-(e)) $y$ ($x$=0.25) for Ni$_{2-x}$(Fe/Co)$_{x}$Mn$_{1+z-y}$Cu$_{y}$Sb$_{1-z}$ systems with different $z$. Closed symbols and open symbols represent results calculated by Monte Carlo Simulation (MCS) and Mean Field Approximation (MFA) methods, respectively.} 
\label{fig5}
\end{figure}

In Fig.~\ref{fig5}, we present the calculated Curie temperature (T$_c^A$) in the austenite phase for all the compositions, using both the Mean-field approximation and the more accurate Monte Carlo simulation method. The results for single-substituted compounds are presented in Fig.~\ref{fig5}(a)-(c), and those for the cosubstituted compounds are presented in Fig.~\ref{fig5}(d)-(e). The T$_c^A$ values for all the cosubstituted compositions are listed in Table \ref{table1}. Here, too, the trends of variations in T$_{c}^{A}$ for cosubstituted compounds can be correlated to the trends in cases of single-substitutions. Overall it can be concluded that for cosubstituted systems, the T$_c^A$ values are higher for compounds with  $z=0.75$. This is more prominent for Co and Cu cosubstituted systems (Co@Ni-Cu@Mn). Thus cosubstituted Co@Ni-Cu@Mn family, with $z=0.75$ have more possibility of fulfilling the target properties of a giant magnetocaloric material.       

\subsection{Prediction of new compounds} \label{prediction}

\begin{table*}[t]
\centering
\caption{\label{table1} Calculated values of equilibrium lattice constant (a$_{0}$), formation energies (E$_{f}$), and total magnetic moment (M$_{\rm A}$) of the systems under considerations in their austenite phases. The corresponding ground state magnetic configurations are shown. The total energy difference ($\Delta$E) between the austenite(L2$_{1}$) and the martensite(tetragonal) phases, the equilibrium value of tetragonal distortion (c$/$a), the differences in magnetic moments between the austenite and martensite phases ($\Delta$M), the Mean-field approximated (T$_c^{A(MFA)}$) and Monte Carlo simulated (T$_c^{A(MCS)}$) Curie temperatures in the austenite phases are also shown. Boldfaces indicate the reference compositions around which experimental studies\cite{22khan2007,21j2007,nayak2009} are available. The compositions shown within the borders are found to satisfy the criteria for considering them as  
efficient giant magnetocaloric materials with improved MCE properties than the reference ones.}
\resizebox{0.88\textwidth}{!}{%
\begin{tabular}{l@{\hspace{0.5cm}} l@{\hspace{0.5cm}} c@{\hspace{0.5cm}} c@{\hspace{0.5cm}} c@{\hspace{0.5cm}} c@{\hspace{0.5cm}} c@{\hspace{0.5cm}} c@{\hspace{0.5cm}} c@{\hspace{0.5cm}} c@{\hspace{0.2cm}} c@{\hspace{0.3cm}} }
\hline\hline
\vspace{-0.33 cm}
\\ \multicolumn{8}{l}{\bf{Ni$_{2-x}$A$_{x}$Mn$_{1.50-y}$Cu$_{y}$Sb$_{0.50}$(z=0.50)}} \\
\hline
\\ \multicolumn{2}{l}{Composition} & Mag. & a$_{0}$ & E$_{f}$ & M$_{\rm A}$ & $\Delta$E & $c/a$ & $\mid$$\Delta$M$\mid$ & T$_c^{A(MFA)}$ & T$_c^{A(MCS)}$ \\
  x & y                            &  Config. & (\r{A}) & (eV$/$f.u.) & ($\mu_{\rm B}/f.u.$) & (meV/atom) &  & ($\mu_{\rm B}/f.u.$) &  (K) & (K) \\ 
\hline
\bf{0.00} & \bf{0.00} & \bf{FIM} & 5.94 & -0.552 & 1.71 & 27.64 & 1.34 & 0.16 & 397 & 422   \\
\hline
\bf{A=Fe}  \\
\hline
0.25 & 0.00 & FIM & 5.92 & -0.450 & 2.09 & 10.68  & 1.29  & 0.18 & 351 & 370   \\
0.25 & 0.25 & FIM & 5.91 & -0.386 & 3.29 & 4.45   & 1.24  & 0.08 & 297 & 340   \\
0.25 & 0.50 & FM  & 5.90 & -0.317 & 4.44 & No MPT &  -    & -    & 327 & 300  \\
\hline
\bf{A=Co}  \\
\hline
\bf{0.25} & \bf{0.00} & \bf{FIM} & 5.93 & -0.670 & 1.99  & 20.97 & 1.30 & 0.04 & 392 & 410      \\
0.25 & 0.25 & FIM & 5.91 & -0.620 & 3.22  & 16.91 & 1.28 & 0.23 & 333 & 401       \\
0.25 & 0.50 & FM  & 5.90 & -0.567 & 4.42  & 10.93 & 1.25 & 0.19 & 355 & 399     \\
\hline
\vspace{0.05 cm}
\\ \multicolumn{7}{l}{\bf{Ni$_{2-x}$A$_{x}$Mn$_{1.75-y}$Cu$_{y}$Sb$_{0.25}$(z=0.75)}} \\
\hline
\\ \multicolumn{2}{l}{Composition} & Mag. & a$_{0}$ & E$_{f}$ & M$_{\rm A}$ & $\Delta$E & $c/a$ & $\mid$$\Delta$M$\mid$ & T$_c^{A(MFA)}$ & T$_c^{A(MCS)}$ \\
  x & y                            &  Config. & (\r{A}) & (eV$/$f.u.) & ($\mu_{\rm B}/f.u.$) & (meV/atom) &  & ($\mu_{\rm B}/f.u.$) &  (K) & (K) \\ 
\hline
0.00 & 0.00 & FIM & 5.86 & -0.470 & 0.69 & 55.32 & 1.36 & 0.15 & 348 & 290  \\
\hline
\bf{A=Fe}  \\
\hline
0.25 & 0.00 & FM                             & 5.89 & -0.352 & 7.71 & 41.22 & 1.38 & 6.27   & 498 & 406    \\
0.25                    & 0.25      & FIM    & 5.84 & -0.312 & 2.17 & 35.30 & 1.35 & 0.19 & 241 & 120  \\
\boxit{1.5in}\bf{0.25} & \bf{0.50} & \bf{FM} & 5.83 & -0.240 & 5.54 & 23.58  & 1.33 & 3.33 & 448 & 362  \\
\boxit{1.5in}\bf{0.25} & \bf{0.75} & \bf{FM} & 5.81 & -0.167 & 4.43 & 24.22  & 1.32 & 1.22 & 408 & 320  \\
\hline
\bf{A=Co}  \\
\hline
\boxit{1.5in}\bf{0.25} & \bf{0.00} & \bf{FM} & 5.90 & -0.615 & 7.75 & 37.25  & 1.34 & 6.78  & 519 & 450  \\
\boxit{1.5in}\bf{0.25} & \bf{0.25} & \bf{FM} & 5.87 & -0.576 & 6.66 & 31.87  & 1.32 & 4.66  & 489 & 460  \\
0.25                    & 0.50      & FM     & 5.84 & -0.525 & 5.62 & 12.20  & 1.28 & 2.53  & 464 & 450  \\
0.25                    & 0.75      & FM     & 5.81 & -0.468 & 4.53 & 4.40   & 1.24 & 0.27  & 431 & 425  \\
\hline
\end{tabular}
}
\end{table*}

Based upon the results presented in the previous three sub-sections, we are now in a position to predict new compounds which can exhibit better MCE properties than that observed in the experimentally synthesized compounds. To this end, we first consider the compounds Ni$_{2}$Mn$_{1.5}$Sb$_{0.5}$ (i.e. x=0.00, y=0.00, z=0.50) and Ni$_{1.75}$Co$_{0.25}$Mn$_{1.5}$Sb$_{0.5}$(i.e. x=0.25, y=0.00, z=0.50) (boldfaced in Table \ref{table1}) as the reference ones with respect to which we assess the improvement in properties. These compounds are chosen as the compositions in these compounds are very close to the experimentally investigated ones\cite{22khan2007,72khan2008,21j2007,nayak2009,anayak2009,60sahoo2014,64han2008}. Comparing all the quantities, we predict four compositions (bold bordered in Table \ref{table1}), two in Fe@Ni-Cu@Mn family, and the other two in Co@Ni-Cu@Mn family. In all the cases, $z=0.75$ that is Mn atom is in excess by 75$\%$ in comparison to perfectly ordered Ni$_{2}$MnSb, the parent compound. For all the predicted compositions, the ground state magnetic configuration in the austenite phase is the FM one, where the two types of Mn atoms are aligned parallel, leading to larger magnetic moments compared to the reference systems.  For these compounds, large values of change in magnetization ($\Delta$M), compared to the reference compounds, are observed during the martensitic phase transformation. Finally, the conditions that the martensitic transformation temperature (T$_{M}$) and Curie temperature (T$_c^A$) either should nearly coincide or T$_c^A$ should be higher than T$_{M}$, are satisfied for the predicted compounds. In order to establish this, we have made an estimation of T$_{M}$ the following way: the value of $\Delta$E corresponding to Ni$_{1.75}$Co$_{0.25}$Mn$_{1.50}$Sb$_{0.50}$ composition is mapped to the experimental martensitic transformation temperature (T$_{M}$) value of 262~K, found for a compound with almost same composition Ni$_{1.8}$Co$_{0.2}$Mn$_{1.5}$Sb$_{0.5}$\cite{nayak2009}. Using this mapping, we found that approximate values of T$_{M}$ are 294~K, 302~K, 465~K and 398~K for Ni$_{1.75}$Fe$_{0.25}$Mn$_{1.25}$Cu$_{0.5}$Sb$_{0.25}$, Ni$_{1.75}$Fe$_{0.25}$MnCu$_{0.75}$Sb$_{0.25}$, Ni$_{1.75}$Co$_{0.25}$Mn$_{1.75}$Sb$_{0.25}$ and Ni$_{1.75}$Co$_{0.25}$Mn$_{1.50}$Cu$_{0.25}$Sb$_{0.25}$, respectively. A look at Table \ref{table1}, along with these mapped values, shows that the above mentioned conditions are satisfied for all four. \\

\subsection{Computation of the MCE parameters} \label{monte-carlo}
After screening the compounds, most suitable to exhibit giant MCE, we aimed at the calculations of the MCE parameters $\Delta$S$_\mathrm{mag}$ and $\Delta$T$_\mathrm{ad}$ to establish our predictions. Since there is no experimental result to compare in cases of the new cosubstituted compounds, it is imperative that our approach of using the DFT, in conjunction with the proposed model Hamiltonian, is validated. To this end, we used our approach to compute the MCE parameters for two compounds: Ni$_{2}$Mn$_{1.52}$Sb$_{0.48}$ and Ni$_{1.8}$Co$_{0.2}$Mn$_{1.52}$Sb$_{0.48}$ where experimental results are available\cite{22khan2007,nayak2009}. After validation, we computed the MCE parameters for Ni$_{1.75}$Co$_{0.25}$Mn$_{1.50}$Cu$_{0.25}$Sb$_{0.25}$, one of the compounds predicted. Due to the huge computational cost involved in cosubstituted compounds with multi-sublattice disorder, we restricted ourselves to only one out of the four new predicted compounds. \\

The compound Ni$_{2}$Mn$_{1.52}$Sb$_{0.48}$ was considered first. The Monte Carlo calculations were done using a simulation domain consisting of 8192 atoms obtained by replicating the unit cell, containing 16 atoms, eight times in each direction. The Mn2 atoms are randomly distributed on the Sb sub-lattices. The final simulation domain contains 983 Sb, 1065 Mn2, 2048 Mn1, 4096 Ni atoms. \\

The magnetic exchange parameters, $J_{ij}$, in equation (\ref{eq-potts}), were obtained from Fig.~1(a) and Fig.~1(b), supplementary material. The magnetic spin states (q) for Ni, Mn2, Mn1 atoms were taken as 3, 6, and 6, respectively, in accordance with the studies on other Heusler systems\cite{1sokolovskiy2013,sokolovskiy2013}. The spin state of each magnetic atom site ($S_{i}$) were chosen randomly by generating a random number between 0 and 1 (0 $\leq$ $r$ $\leq$ 1) and selecting the state as: if 0 $\leq$ $r$ $\leq$ l/3, then q$_\mathrm{Ni}$ = l, l = 1, 2, 3, and if 0 $\leq$ $r$ $\leq$ k/6, then q$_\mathrm{Mn1(Mn2)}$ = k, k = 1, 2, 3, . . . 6. We considered lattice sites up to the third coordination shells for Mn1-Mn1, Mn2-Mn2, Mn1-Mn2 pairs, and up to the first-coordination shell for Mn1(Mn2)-Ni and Ni-Ni atom pairs in the summation.
For elastic part of the Hamiltonian (equation (\ref{eq-struc})), the summation was taken over the pairs up to the second-coordination shell. A similar procedure, as for choosing $S_{i}$ values, had been used to assign the $\sigma_i$ values randomly. Values of structural constants $J$ and $K$ were chosen such that the martensitic transformation temperature (T$_{M}$) could be adjusted around the experimental T$_{M}$, which is around the room temperature ($\approx$300~K)\cite{22khan2007,72khan2008}. The constraint of $K/J\leq 0.5$ was imposed to get rid of any pre-martensitic phase. Although $J$ and $K$ could be obtained from \textit{ab-initio} calculations, we used a simple procedure of tuning to reduce the complexity as well as computational cost. This had been followed in other investigations\cite{buchelnikov2010,singh2013} too. The degeneracy factor ($p$) for the cubic phase was taken as 2, since the cubic phase can distort along one of the three directions (here along the z-direction). The $K1$ was chosen to be negative since it was experimentally observed that T$_{M}$ decreased under application of external magnetic field. The negative  $K1$ fixed the ghost deformation state $\sigma_{g}$ to -1. The value of $K1$ was chosen such that the maximum magnetic entropy change is obtained around the experimental martensitic transformation temperature.   \\
In the magneto-elastic interaction part (equation (\ref{eq-interaction})), the interaction parameters $U_{ij}$ in cubic ($U_c$) and tetragonal ($U_t$) phases were tuned in such a way that the Curie temperature in the austenite phase is obtained around the experimental Curie temperature (350~K for the compound considered). 

Thus, with an initial guess of $J$ and $K$ values, we adjusted the T$_{M}$, to bring it closer to the experimental value. Once a reasonable T$_{M}$ is obtained, we tuned the parameter U$_{ij}$ in both structural phases to obtain the experimental Curie temperature. These are done by adjusting the co-ordination shells over which summations are done. These were done in the absence of an external magnetic field, i.e., H$_\mathrm{ext}$=0. Then we applied an external field of 5~T and tuned the $K1$ parameter so that T$_{M}$ shifts in the direction observed in the experiment. This ensured a correct behavior of magnetic entropy change $\Delta$S$_\mathrm{mag}$  as a function of temperature and $\Delta$S$_\mathrm{mag}^{max}$ is achieved around the experimental T$_{M}$. Here, depending on the sign of K1, ghost deformation state $\sigma_{g}$ was chosen. The simulation started with the initial values of $\sigma_{i}$ as 1 for all the lattice sites and $S_{i}$, $S_{g}$ as 1 for the magnetic atoms. The final values of all the parameters are presented in Table \ref{parameters}.  For each temperature step, 2$\times$10$^5$ MC steps were performed. The system was equilibrated for 1$\times$10$^5$ MC steps, and the data, then, were collected for 10$^5$ steps. For each temperature step, energy of the system ($H$), magnetization ($m$), and structural distortion ($\varepsilon$) were averaged over 1000 data points collected after every 100 MC steps. These averaged quantities were then used to calculate the various thermodynamic quantities using equations, given in Section \ref{2-method}. \\

\begin{table}[t]
\centering
\caption{\label{parameters} Values of structural, magneto-elastic and coupling parameters that are used in the present study for Ni$_{2-x}$Co$_{x}$Mn$_{1+z-y}$Cu$_{y}$Sb$_{1-z}$ compounds. }
\resizebox{0.49\textwidth}{!}{%
\begin{tabular}{c@{\hspace{0.4cm}} c@{\hspace{0.4cm}} c@{\hspace{0.8cm}} c@{\hspace{0.4cm}} c@{\hspace{0.4cm}} c@{\hspace{0.4cm}} c@{\hspace{0.4cm}} c@{\hspace{0.2cm}} }
\hline\hline
\vspace{-0.33 cm}
\\ \multicolumn{3}{c}{Concentrations} & \multicolumn{5}{c}{Parameters} \\
\hline
  $x$  &   $y$   & $z$                     & $J$   & $K$   & $U_{c}$ & $U_{t}$ & $K1$ \\
     &       &                       & meV & meV & meV     & meV     &     \\
\hline
0.00 & 0.00  & 0.52                  &  1.67   &  0.25   & 1.36   &  3.54        &  -0.25   \\
0.20 & 0.00  & 0.52                  &  1.67   &  0.42   & 7.56   &  14.24       & -0.8     \\
0.25 & 0.25  & 0.75                  &  2.00   &  0.32   & 7.56   &  14.24       & -0.4     \\
\hline

\hline
\end{tabular}
}
\end{table}

\begin{figure*}[t]
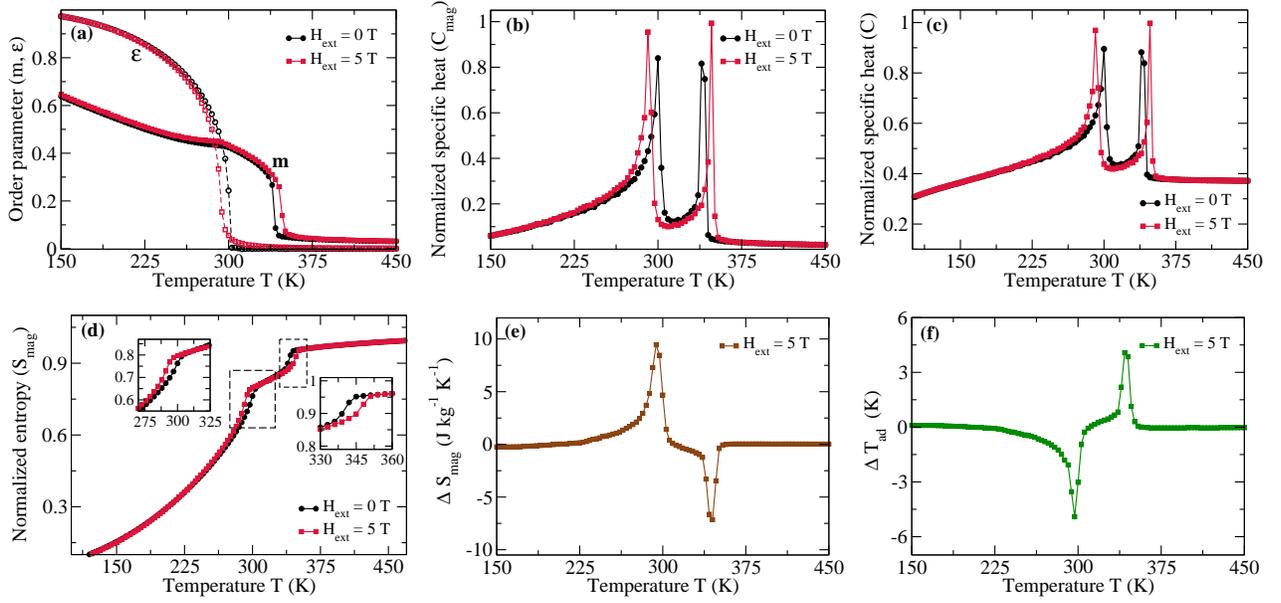

\centerline{\hfill
\psfig{file=fig6a.eps,width=0.3\textwidth}
\hspace{0.005cm}
\psfig{file=fig6b.eps,width=0.305\textwidth}
\hspace{0.005cm}
\psfig{file=fig6c.eps,width=0.3\textwidth}
\hfill}
\vspace{0.2cm}
\centerline{\hfill
\psfig{file=fig6d.eps,width=0.295\textwidth}
\hspace{0.005cm}
\psfig{file=fig6e.eps,width=0.31\textwidth}
\hspace{0.005cm}
\psfig{file=fig6f.eps,width=0.295\textwidth}
\hfill}
\caption{The calculated temperature dependence of (a) magnetic ($m$) and strain ($\varepsilon$) order parameters, (b)-(c) normalized magnetic and total specific heat ($C_\mathrm{mag}$ and $C$, respectively), (d) normalized magnetic entropy (S$_\mathrm{mag}$) in an external field of 0 and 5~T and the temperature dependence of (e) the isothermal magnetic entropy change ($\Delta$S$_\mathrm{mag}$) and (f) adiabatic temperature change ($\Delta$T$_\mathrm{ad}$) due to change in the magnetic field from 0 to 5~T are shown. The results are for Ni$_{2-x}$(Fe/Co)$_{x}$Mn$_{1+z-y}$Cu$_{y}$Sb$_{1-z}$, with $z=0.52, x=0.00, y=0.00$.}
\label{fig6}
\end{figure*}

Under zero field, the strain order parameter ($\varepsilon$) shows the structural transformation from austenite (undistorted phase with $\varepsilon=0$) to martensite phases (distorted phase with $\varepsilon=1$) with decreasing temperature (Fig.~\ref{fig6}(a)). The transformation occurs around 300~K, which is in a good agreement with the experimental T$_{M}$. The magnetic order parameter ($m$) is almost zero at high temperatures indicating a paramagnetic phase. With a decrease in temperature, the magnetic order parameter increases gradually, indicating the transformation from the paramagnetic to a ferromagnetic phase around 350~K. Thus, the magnetic transition temperature (T$_c^A$) in the austenite phase also matches very well with the experimental value. With a further decrease in temperature, near the T$_{M}$, a small kink, indicating a weak magneto-elastic coupling, is observed in the magnetic order parameter ($m$). With an applied external magnetic field of 5~T, T$_{M}$ decreases, and the T$_c^A$ increases in agreement with the experimental trend. \\

The calculated magnetic specific heat ($C_\mathrm{mag}$) (equation \ref{eq-cmag}) is shown in Fig.~\ref{fig6}(b). The total specific heat is also calculated as a function of temperature by calculating the lattice specific heat ($C_\mathrm{lat}$) (equation \ref{eq-clat}). In absence of experimental result on this compound, the Debye temperature $\Theta_{D}$ was taken as 222~K, the experimental $\Theta_{D}$ of Ni$_{2}$MnSb\cite{podgornykh2007}. Here, we assumed that the lattice specific heat does not contribute significantly to the isothermal entropy change, i.e., there is no significant impact of the application of magnetic field on $C_\mathrm{lat}$. The isothermal entropy change, from lattice contributions across the magneto-structural transition, is not significant as long as the Debye temperature does not depend strongly on the magnetization and magnetic field. Two peaks can be observed in the specific heat curves, one at higher temperature corresponds to the second-order magnetic transition from paramagnetic to ferromagnetic phase, while the other at lower temperature corresponds to the structural transformation from austenite to martensite phases. \\

The magnetic entropy curve (in Fig.~\ref{fig6}(d)) has been obtained by integrating the magnetic specific heat curves using equation \ref{eq-smag} both at zero field and a field of 5~T. At very low temperatures, the calculated entropy has lower values, as expected, and increases with an increase in temperature, saturating at high temperatures beyond the magnetic transformation in the austenite phase. Upon application of the external magnetic field, the entropy of the system decreases as the system undergoes the magnetic transformation, while the entropy increases at the structural transformation. The insets of Fig.~\ref{fig6}(d) show the changes in the entropy of the system when the structural (inset with lower temperature range) and magnetic (inset with higher temperature range) transformations take place. \\

The magnetic field induced isothermal entropy change, $\Delta$S$_{\text {mag}}$ (equation \ref{eq-del-smag}) and the adiabatic temperature change $\Delta$T$_\mathrm{ad}$ (equation \ref{eq-del-Tad}) are shown in Fig.~\ref{fig6}(e) and Fig.~\ref{fig6}(f) respectively. The maximum change in entropy is positive for structural transformation and negative for the magnetic transformation. Hence, we have an inverse magnetocaloric effect (cooling of the material in the presence of magnetic field) during structural transformation, while regular magnetocaloric effect (heating of the material in the presence of magnetic field) as the magnetic transformation takes place. A maximum value of isothermal entropy change of 9.8 Jkg$^{-1}$K$^{-1}$ is obtained at the first order magneto-structural transition, in good agreement with the experimental observation\cite{22khan2007,nayak2009}. Our calculations predict a large value of nearly 5~K of $\Delta$T$_{\text ad}$ which, however, could not be compared with experiments due to the unavailability.  

\begin{figure}[h]
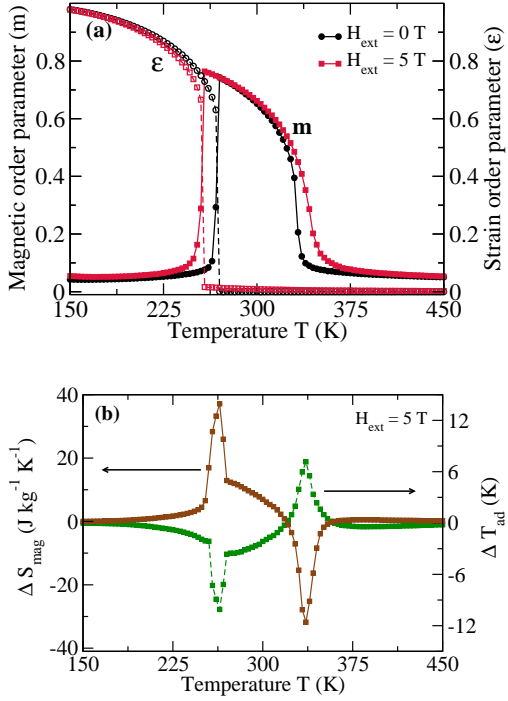

\centerline{\hfill
\psfig{file=fig7a.eps,width=0.37\textwidth}
\hfill}
\vspace{0.5cm}
\centerline{\hfill
\psfig{file=fig7b.eps,width=0.37\textwidth}
\hfill}
\caption{The calculated temperature dependence of (a) magnetic ($m$) and strain ($\varepsilon$) order parameter in an external field of 0 and 5~T and (b)  the isothermal magnetic entropy change ($\Delta$S$_\mathrm{mag}$) and adiabatic temperature change ($\Delta$T$_\mathrm{ad}$) due to change in the magnetic field from 0 to 5~T in Ni$_{2-x}$Co$_{x}$Mn$_{1+z-y}$Cu$_{y}$Sb$_{1-z}$ ; $z=0.52, x=0.20, y=0.00$.}
\label{fig7}
\end{figure}

We next applied the same formalism to the Co substituted compound Ni$_{1.8}$Co$_{0.2}$Mn$_{1.52}$Sb$_{0.48}$. Due to the availability of experimental results \cite{nayak2009}, we could make a direct comparison. The results are shown in Fig.~\ref{fig7}. The {\it ab initio} magnetic exchange parameters used here are shown in  Fig.~1(c)-1(d), supplementary material. In here, the number of spin states for Co was taken to be q$_\mathrm{Co}$=4. The parameters in the Hamiltonian were adjusted, such that the experimental value of T$_{M}$ ($\approx$ 260~K) and T$_c^A$ ($\approx$ 330~K) could be reproduced. This is evident from the curves of $m$ and $\varepsilon$ in Fig.~\ref{fig7}(a).

\begin{figure}[t]
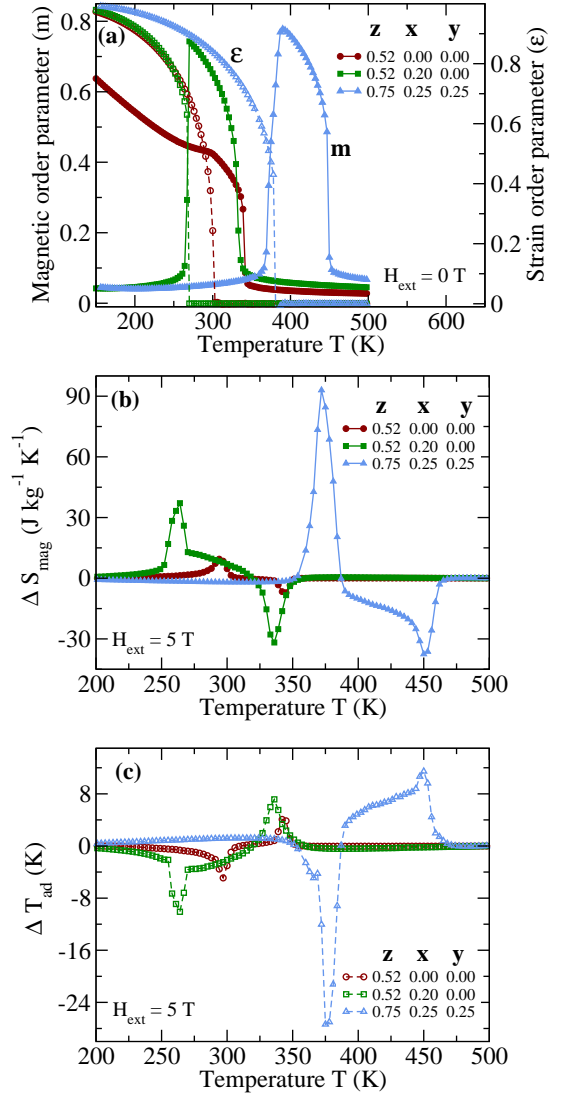

\centerline{\hfill
\psfig{file=fig8a.eps,width=0.36\textwidth}
\hfill}
\vspace{0.3cm}
\centerline{\hfill
\psfig{file=fig8b.eps,width=0.36\textwidth}
\hfill}
\vspace{0.3cm}
\centerline{\hfill
\psfig{file=fig8c.eps,width=0.36\textwidth}
\hfill}
\caption{The calculated temperature dependence of (a) magnetic ($m$) and strain ($\varepsilon$) order parameter in absence of an external field, (b) the isothermal magnetic entropy change ($\Delta$S$_\mathrm{mag}$) and (c) adiabatic temperature change ($\Delta$T$_\mathrm{ad}$) due to change in the magnetic field from 0 to 5~T in Ni$_{2-x}$Co$_{x}$Mn$_{1+z-y}$Cu$_{y}$Sb$_{1-z}$; $z=0.75, x=0.25, y=0.25$. Results for the reference compositions, Ni$_{2}$Mn$_{1.52}$Sb$_{0.48}$ and Ni$_{1.8}$Co$_{0.2}$Mn$_{1.52}$Sb$_{0.48}$ have also been included for comparison.}  
\label{fig8}
\end{figure}

The parameters used for this calculation are listed in Table \ref{parameters}. The directions of shifts in T$_{M}$ and T$_c^A$ under application of an external magnetic field of 5~T are in agreement with the experimental trends. One noteworthy point is that in contrast to the Ni$_{2}$Mn$_{1.52}$Sb$_{0.48}$, magnetization changes sharply at the structural transition in this case. This is important to obtain a giant MCE. This significant change in magnetization can be understood by analyzing the magnetic exchange interactions between different atom pairs in both structural phases (Fig.~1(c) and Fig.~1(d), supplementary material). While for both Ni$_{2}$Mn$_{1.52}$Sb$_{0.48}$ and Ni$_{1.8}$Co$_{0.2}$Mn$_{1.52}$Sb$_{0.48}$, the dominant antiferromagnetic interaction between Mn atoms are four times larger in the tetragonal phase, in comparison to that in the austenite phase, larger ferromagnetic interactions due to the Mn1(Mn2)-Co atom pairs in the austenite phase appears for the later compound. This can be correlated to the larger change in the magnetization for Co substituted compound. The MCE quantities $\Delta$S$_\mathrm{mag}$ and $\Delta$T$_\mathrm{ad}$, calculated as a function of temperature, as shown in Fig.~\ref{fig7}(b) and the maximum $\Delta$S$_\mathrm{mag}$ of nearly 40~Jkg$^{-1}$K$^{-1}$, much larger than the Ni$_{2}$Mn$_{1.52}$Sb$_{0.48}$, is obtained due to the magneto-structural transition near T$_{M}$ in an applied field of 5~T. This is in excellent agreement with the experimental value\cite{nayak2009} of 35~Jkg$^{-1}$K$^{-1}$. We also obtained a large value of $\Delta$T$_{\text ad}$ that could not be verified in absence of experimental results. \\

The excellent agreement of the results obtained for the two compounds with the experimental observations validates the approach adopted here for computing the variables quantifying MCE. Therefore, we apply the same formalism for the cosubstituted compound Ni$_{1.75}$Co$_{0.25}$Mn$_{1.50}$Cu$_{0.25}$Sb$_{0.25}$. The results are presented in Fig.~\ref{fig8}. The \textit{ab initio} calculated exchange interactions used are shown in Fig.~1(e) and Fig.~1(f), supplementary material. The values of elastic and magneto-elastic parameters were tuned (Table~\ref{parameters}) to fix the T$_{M}$ at 398~K, as predicted in subsection \ref{prediction}, and T$_c^A$ at 460~K as calculated through MCS in Table~\ref{table1}. For the purpose of comparison, we have included the results for Ni$_{2}$Mn$_{1.52}$Sb$_{0.48}$ and Ni$_{1.8}$Co$_{0.2}$Mn$_{1.52}$Sb$_{0.48}$. We find that cosubstitution leads to an increase in the working temperature (T$_{M}$). Also, the change in magnetization near MPT is larger (Fig.~\ref{fig8}(a)). Both these characteristics were desired for a larger MCE in cosubstituted compounds. The calculated MCE quantities (Fig.~\ref{fig8}(b) and \ref{fig8}(c) meet this expectation. The results demonstrate that $\Delta$S$_\mathrm{mag}$ four times higher than Ni$_{2}$Mn$_{1.52}$Sb$_{0.48}$ and two times higher than the Ni$_{1.8}$Co$_{0.2}$Mn$_{1.52}$Sb$_{0.48}$ are obtained. Stronger ferromagnetic interactions in the austenite phase of cosubstituted compound (Fig.~1(e)-(f), supplementary material), in comparison with the other two compounds, can be correlated to this amplified effect.      

\section{Conclusions}
Using first-principles electronic structure calculations, we provide a protocol to systematically screen materials, potential to exhibit giant MCE driven by a first-order magneto-structural transition at temperatures near room temperature or above, among given Heusler family of compounds. We apply this approach to find target compounds in cosubstituted Ni-Mn-Sb family; the cosubstitution done at Ni and Mn sites. Our approach predicted four new compounds in the two cosubstituted families. In order to validate our predictions, we took recourse to a thermodynamic model to compute the MCE properties in one of these predicted compounds. The robustness and accuracy of the computational approach using the thermodynamic model that takes into account magnetic, elastic and magneto-elastic effects in equal footing, is demonstrated by computing the MCE parameters in Ni$_{2}$Mn$_{1.52}$Sb$_{0.48}$ and Co substituted Ni$_{1.8}$Co$_{0.2}$Mn$_{1.52}$Sb$_{0.48}$ compounds where experimental results are available. Further computation of MCE parameters for one of the predicted compounds yields magnetic entropy change as large as four times in comparison to that observed experimentally. Thus, this established the protocol for screening materials from a large database adopted in this work. This work, apart from demonstrating the power of {\it ab initio} based approaches for computations of MCE parameters,  offers experimentalists a broader scope to explore new materials where giant MCE, driven by first-order magneto-elastic transition, can be realized through cosubstitution in Heusler compounds. 

\section{Acknowledgement}
The authors gratefully acknowledge the Department of Science and Technology, India, for the computational facilities under Grant No. SR/FST/P-II/020/2009 and IIT Guwahati for the PARAM supercomputing facility.
\bibliographystyle{aip} 

\begin{thebibliography}{10}

\bibitem{oliva1989}
L.~P. Oliva,
\newblock Pace Envtl. L. Rev. {\bf 7}, 213 (1989).

\bibitem{tegusi2003}
T.~Tegusi et~al.,
\newblock {\em Novel materials for magnetic refrigeration},
\newblock Universiteit van Amsterdam [Host], 2003.

\bibitem{banks2002}
J.~Banks,
\newblock 2002 Report of the Methyl Bromide Technical Options Committee, UNEP
  (2002).

\bibitem{hughes2007}
I.~Hughes et~al.,
\newblock Nature {\bf 446}, 650 (2007).

\bibitem{pecharsky1997}
V.~K. Pecharsky and K.~A. Gschneidner~Jr,
\newblock Phys. Rev. Lett. {\bf 78}, 4494 (1997).

\bibitem{pecharsky2001}
V.~K. Pecharsky and K.~A. Gschneidner~Jr,
\newblock Advanced Materials {\bf 13}, 683 (2001).

\bibitem{tishin2016}
A.~M. Tishin and Y.~I. Spichkin,
\newblock {\em The magnetocaloric effect and its applications},
\newblock CRC Press, 2016.

\bibitem{de2006}
A.~De~Campos et~al.,
\newblock Nature materials {\bf 5}, 802 (2006).

\bibitem{fujieda2002}
S.~Fujieda, A.~Fujita, and K.~Fukamichi,
\newblock Appl. Phys. Lett. {\bf 81}, 1276 (2002).

\bibitem{wada2001}
H.~Wada and Y.~Tanabe,
\newblock Appl. Phys. Lett. {\bf 79}, 3302 (2001).

\bibitem{tegus2002}
O.~Tegus, E.~Br{\"u}ck, and L.~Zhang,
\newblock Physica B {\bf 319}, 174 (2002).

\bibitem{li2012}
Z.~Li, J.~L. S{\'a}nchez~Llamazares, C.~F. S{\'a}nchez-Vald{\'e}s, Y.~Zhang,
  C.~Esling, X.~Zhao, and L.~Zuo,
\newblock Appl. Phys. Lett. {\bf 100}, 174102 (2012).

\bibitem{li2014}
Z.~Li, Y.~Zhang, C.~F. S{\'a}nchez-Vald{\'e}s, J.~L. S{\'a}nchez~Llamazares,
  C.~Esling, X.~Zhao, and L.~Zuo,
\newblock Appl. Phys. Lett. {\bf 104}, 044101 (2014).

\bibitem{krenke2005}
T.~Krenke, M.~Acet, E.~F. Wassermann, X.~Moya, L.~Ma{\~n}osa, and A.~Planes,
\newblock Phys. Rev. B {\bf 72}, 014412 (2005).

\bibitem{7krenke2007}
T.~Krenke et~al.,
\newblock Phys. Rev. B {\bf 75}, 104414 (2007).

\bibitem{8pasquale2005}
M.~Pasquale, C.~P. Sasso, L.~H. Lewis, L.~Giudici, T.~Lograsso, and
  D.~Schlagel,
\newblock Phys. Rev. B {\bf 72}, 094435 (2005).

\bibitem{pathak2007}
A.~K. Pathak, M.~Khan, I.~Dubenko, S.~Stadler, and N.~Ali,
\newblock Appl. Phys. Lett. {\bf 90}, 262504 (2007).

\bibitem{79krenke2005}
T.~Krenke, E.~Duman, M.~Acet, E.~F. Wassermann, X.~Moya, L.~Ma{\~n}osa, and
  A.~Planes,
\newblock Nature materials {\bf 4}, 450 (2005).

\bibitem{muthu2010}
S.~E. Muthu, N.~R. Rao, M.~M. Raja, D.~R. Kumar, D.~M. Radheep, and
  S.~Arumugam,
\newblock Journal of Physics D: Applied Physics {\bf 43}, 425002 (2010).

\bibitem{91duc2012}
N.~Duc, T.~Thanh, N.~Yen, P.~Thanh, N.~Dan, and T.~Phan,
\newblock Journal of the Korean Physical Society {\bf 60}, 454 (2012).

\bibitem{22khan2007}
M.~Khan, N.~Ali, and S.~Stadler,
\newblock J. Appl. Phys. {\bf 101}, 053919 (2007).

\bibitem{nayak2009}
A.~K. Nayak, K.~Suresh, and A.~Nigam,
\newblock Journal of Physics D: Applied Physics {\bf 42}, 035009 (2009).

\bibitem{anayak2009}
A.~K. Nayak, K.~Suresh, and A.~Nigam,
\newblock Journal of Physics D: Applied Physics {\bf 42}, 115004 (2009).

\bibitem{64han2008}
Z.~Han, D.~Wang, C.~Zhang, H.~Xuan, J.~Zhang, B.~Gu, and Y.~Du,
\newblock J. Appl. Phys. {\bf 104}, 053906 (2008).

\bibitem{59sahoo2011}
R.~Sahoo, A.~K. Nayak, K.~Suresh, and A.~Nigam,
\newblock J. Appl. Phys. {\bf 109}, 123904 (2011).

\bibitem{zeleny2014}
M.~Zelen{\`y}, A.~Sozinov, L.~Straka, T.~Bj{\"o}rkman, and R.~M. Nieminen,
\newblock Phys. Rev. B {\bf 89}, 184103 (2014).

\bibitem{55sokolovskiy2015}
V.~V. Sokolovskiy, P.~Entel, V.~Buchelnikov, and M.~Gruner,
\newblock Phys. Rev. B {\bf 91}, 220409 (2015).

\bibitem{perez2018}
A.~Perez-Checa, J.~Feuchtwanger, J.~Barandiaran, A.~Sozinov, K.~Ullakko, and
  V.~Chernenko,
\newblock Scripta Materialia {\bf 154}, 131 (2018).

\bibitem{ghosh2020co}
S.~Ghosh and S.~Ghosh,
\newblock Physical Review Materials {\bf 4}, 025401 (2020).

\bibitem{amaral2007}
J.~Amaral, N.~Silva, and V.~Amaral,
\newblock Applied Physics Letters {\bf 91}, 172503 (2007).

\bibitem{alvarez2011}
P.~{\'A}lvarez, P.~Gorria, and J.~Blanco,
\newblock Physical Review B {\bf 84}, 024412 (2011).

\bibitem{szalowski2011}
K.~Sza{\L}owski, T.~Balcerzak, and A.~Bob{\'a}k,
\newblock Journal of magnetism and magnetic materials {\bf 323}, 2095 (2011).

\bibitem{triguero2006}
C.~Triguero, M.~Porta, and A.~Planes,
\newblock Physical Review B {\bf 73}, 054401 (2006).

\bibitem{nobrega2006}
E.~Nobrega, N.~de~Oliveira, P.~Von~Ranke, and A.~Troper,
\newblock Journal of Physics: Condensed Matter {\bf 18}, 1275 (2006).

\bibitem{1buchelnikov2010}
V.~D. Buchelnikov, V.~Sokolovskiy, S.~Taskaev, and P.~Entel,
\newblock Theoretical modeling of magnetocaloric effect in heusler ni-mn-in
  alloy by monte carlo study,
\newblock in {\em Materials Science Forum}, volume 635, pages 137--142, Trans
  Tech Publ, 2010.

\bibitem{buchelnikov2011}
V.~Buchelnikov et~al.,
\newblock Journal of Physics D: Applied Physics {\bf 44}, 064012 (2011).

\bibitem{buchelnikov2010}
V.~Buchelnikov et~al.,
\newblock Phys. Rev. B {\bf 81}, 094411 (2010).

\bibitem{nobrega2005}
E.~N{\'o}brega, N.~de~Oliveira, P.~von Ranke, and A.~Troper,
\newblock Physical Review B {\bf 72}, 134426 (2005).

\bibitem{singh2013}
N.~Singh and R.~Arr{\'o}yave,
\newblock J. Appl. Phys. {\bf 113}, 183904 (2013).

\bibitem{1sokolovskiy2013}
V.~Sokolovskiy, V.~Buchelnikov, S.~Taskaev, V.~Khovaylo, M.~Ogura, and
  P.~Entel,
\newblock Journal of Physics D: Applied Physics {\bf 46}, 305003 (2013).

\bibitem{sokolovskiy2013}
V.~Sokolovskiy et~al.,
\newblock J. Appl. Phys. {\bf 114}, 183913 (2013).

\bibitem{41blochl1994}
P.~E. Bl{\"o}chl,
\newblock Phys. Rev. B {\bf 50}, 17953 (1994).

\bibitem{43kresse1999}
G.~Kresse and D.~Joubert,
\newblock Phys. Rev. B {\bf 59}, 1758 (1999).

\bibitem{42kresse1996}
G.~Kresse and J.~Furthm{\"u}ller,
\newblock Phys. Rev. B {\bf 54}, 11169 (1996).

\bibitem{44perdew1996}
J.~P. Perdew, K.~Burke, and M.~Ernzerhof,
\newblock Phys. Rev. Lett. {\bf 77}, 3865 (1996).

\bibitem{ebert2011}
H.~Ebert, D.~Koedderitzsch, and J.~Minar,
\newblock Reports on Progress in Physics {\bf 74}, 096501 (2011).

\bibitem{liechtenstein1987}
A.~I. Liechtenstein, M.~Katsnelson, V.~Antropov, and V.~Gubanov,
\newblock Journal of Magnetism and Magnetic Materials {\bf 67}, 65 (1987).

\bibitem{26sokolovskiy2012}
V.~Sokolovskiy, V.~Buchelnikov, M.~Zagrebin, P.~Entel, S.~Sahoo, and M.~Ogura,
\newblock Phys. Rev. B {\bf 86}, 134418 (2012).

\bibitem{meinert2010}
M.~Meinert, J.-M. Schmalhorst, and G.~Reiss,
\newblock J. Phys.: Condens. Matter {\bf 23}, 036001 (2010).

\bibitem{landau2014}
D.~P. Landau and K.~Binder,
\newblock {\em A guide to Monte Carlo simulations in statistical physics},
\newblock Cambridge university press, 2014.

\bibitem{zagrebin2016}
M.~Zagrebin, V.~Sokolovskiy, and V.~Buchelnikov,
\newblock Journal of Physics D: Applied Physics {\bf 49}, 355004 (2016).

\bibitem{28bkundu2017}
A.~Kundu, S.~Ghosh, and S.~Ghosh,
\newblock Phys. Rev. B {\bf 96}, 174107 (2017).

\bibitem{meyer2000}
P.~Meyer,
\newblock School of Mathematics and Computing, University of Derby  (2000).

\bibitem{singh2011}
N.~Singh, E.~Dogan, I.~Karaman, and R.~Arr{\'o}yave,
\newblock Phys. Rev. B {\bf 84}, 184201 (2011).

\bibitem{blume1971}
M.~Blume, V.~Emery, and R.~B. Griffiths,
\newblock Physical review A {\bf 4}, 1071 (1971).

\bibitem{vives1996}
E.~Vives, T.~Cast{\'a}n, and P.-A. Lindg{\aa}rd,
\newblock Physical Review B {\bf 53}, 8915 (1996).

\bibitem{newman1999}
M.~Newman and G.~Barkema,
\newblock {\em Monte carlo methods in statistical physics chapter 1-4},
\newblock Oxford University Press: New York, USA, 1999.

\bibitem{ghosh2020}
S.~Ghosh and S.~Ghosh,
\newblock Physical Review B {\bf 101}, 024109 (2020).

\bibitem{82ghosh2014}
A.~Ghosh and K.~Mandal,
\newblock Appl. Phys. Lett. {\bf 104}, 031905 (2014).

\bibitem{35sanchez2007}
V.~S{\'a}nchez-Alarcos, V.~Recarte, J.~P{\'e}rez-Landaz{\'a}bal, and G.~Cuello,
\newblock Acta Materialia {\bf 55}, 3883 (2007).

\bibitem{ghosh2019}
S.~Ghosh and S.~Ghosh,
\newblock Phys. Rev. B {\bf 99}, 064112 (2019).

\bibitem{45li2011}
C.-M. Li, H.-B. Luo, Q.-M. Hu, R.~Yang, B.~Johansson, and L.~Vitos,
\newblock Phys. Rev. B {\bf 84}, 024206 (2011).

\bibitem{89chakrabarti2013}
A.~Chakrabarti, M.~Siewert, T.~Roy, K.~Mondal, A.~Banerjee, M.~E. Gruner, and
  P.~Entel,
\newblock Phys. Rev. B {\bf 88}, 174116 (2013).

\bibitem{81sokolovskiy2017}
V.~Sokolovskiy, M.~Zagrebin, and V.~Buchelnikov,
\newblock Journal of Physics D: Applied Physics {\bf 50}, 195001 (2017).

\bibitem{21j2007}
J.~Du, Q.~Zheng, W.~J. Ren, W.~J. Feng, X.~G. Liu, and Z.~D. Zhang,
\newblock Journal of Physics D: Applied Physics {\bf 40}, 5523 (2007).

\bibitem{72khan2008}
M.~Khan, I.~Dubenko, S.~Stadler, and N.~Ali,
\newblock J. Phys.: Condens. Matter {\bf 20}, 235204 (2008).

\bibitem{60sahoo2014}
R.~Sahoo, K.~Suresh, and A.~Das,
\newblock Journal of Magnetism and Magnetic Materials {\bf 371}, 94 (2014).

\bibitem{podgornykh2007}
S.~Podgornykh, S.~Streltsov, V.~Kazantsev, and E.~Shreder,
\newblock Journal of Magnetism and Magnetic materials {\bf 311}, 530 (2007).

\end{thebibliography}

\end{document}